\documentclass[preprint,prd,aps,eqsecnum,nofootinbib]{revtex4}
\usepackage{amsmath}

\begin{document}

\def\cstok#1{\leavevmode\thinspace\hbox{\vrule\vtop{\vbox{\hrule\kern1pt
\hbox{\vphantom{\tt/}\thinspace{\tt#1}\thinspace}}
\kern1pt\hrule}\vrule}\thinspace}

\begin{center}
\bibliographystyle{article}
{\Large \textsc{Energy-momentum tensor of a Casimir apparatus in
a weak gravitational field: scalar case}}
\end{center}

\author{Giampiero Esposito$^{1}$ \thanks{Electronic address:
giampiero.esposito@na.infn.it}, 
George M. Napolitano$^{2}$ \thanks{Electronic address:
napolita@na.infn.it},
Luigi Rosa$^{2,1}$ \thanks{Electronic address: 
luigi.rosa@na.infn.it}} 

\affiliation{
${\ }^{1}$Istituto Nazionale di Fisica Nucleare, Sezione di Napoli,\\
Complesso Universitario di Monte S. Angelo, Via Cintia, Edificio 6, 80126
Napoli, Italy\\
${\ }^{2}$Dipartimento di Scienze Fisiche, Complesso Universitario di Monte
S. Angelo,\\
Via Cintia, Edificio 6, 80126 Napoli, Italy}

\vspace{0.4cm}
\date{\today}

\begin{abstract}
Recent work in the literature had evaluated the energy-momentum tensor
of a Casimir apparatus in a weak gravitational field, for an electromagnetic
field subject to perfect conductor boundary conditions on parallel plates.
The Casimir apparatus was then predicted to experience a tiny push in the
upwards direction, and the regularized energy-momentum tensor was found
to have a trace anomaly. The latter, unexpected property made it compelling
to assess what happens in a simpler case. For this purpose, the present
paper studies a free, real massless scalar field subject to homogeneous
Dirichlet conditions on the parallel plates. Working to first order in the
constant gravity acceleration, the resulting regularized and renormalized
energy-momentum tensor is found to be covariantly conserved, while the trace
anomaly vanishes if the massless scalar field is conformally coupled
to gravity. Conformal coupling also ensures a finite Casimir energy and
finite values of the pressure upon parallel plates.
\end{abstract}

\maketitle
\bigskip
\vspace{2cm}

\section{Introduction}

Ever since Casimir discovered that suitable differences of zero-point
energies of the quantized electromagnetic field can be made finite and 
produce measurable effects \cite{KNAWA-51-793}, 
several efforts have been produced to
understand the physical implications and applications of 
this property \cite{PRPLC-353-1, JPAGB-37-R209, RNCIB-27N6-1,
RPPHA-68-201, IJSQE-13-400}
In particular, we are here concerned with the recent theoretical
discovery that Casimir energy gravitates \cite{PHRVA-D74-085011,
PHRVA-D76-025004, JPAGB-40-10935, PHRVA-D76-025008}. 
In Ref. \cite{PHRVA-D74-085011}, some of us
proved this as part of an investigation which led to the evaluation of
the energy-momentum tensor of a Casimir apparatus in a weak gravitational 
field. In that investigation, the functional integral quantization of
Maxwell theory was applied, with perfect conductor boundary conditions
on parallel plates at distance $a$ from each other. On using Fermi--Walker
coordinates, where the $(x_{1},x_{2})$ coordinates span the plates,
while the $z \equiv x_{3}$ axis coincides with the vertical upwards
direction (so that the plates have equations $z=0$ and $z=a$, respectively),
and working to first order in the constant gravity
acceleration $g$, the spacetime metric reads as \cite{PHRVA-D74-085011}
\begin{equation}
ds^{2}=-c^{2} \left(1+\varepsilon {z\over a}\right)dt^{2}
+dx_{1}^{2}+dx_{2}^{2}+dz^{2}+{\rm O}(|x|^{2}),
\label{(1.1)}
\end{equation}
where $\varepsilon \equiv {2ga \over c^{2}}$. 
The resulting regularized
(with point-split method) and renormalized energy-momentum tensor 
$\langle T^{\mu \nu} \rangle$ was found to be covariantly conserved,
with photon and ghost Green functions obeying the mixed boundary
conditions of the problem and satisfying the Ward identities 
\cite{PHRVA-D74-085011}. All
of this would have been completely reassuring, had it not been for the
fact that such a $\langle T^{\mu \nu} \rangle$ was found to have a
trace anomaly
\begin{equation}
\langle T_{\mu}^{\; \mu} \rangle=a^{-3}f \left({z\over a}\right),
\label{(1.2)}
\end{equation}
where
\begin{equation}
f \left({z\over a}\right) \equiv {\pi \over 360} {hg \over c}
\left({z\over a}
-{15\over 2\pi} {\cos (\pi z/a)\over \sin^{3}(\pi z/a)} \right).
\label{(1.3)}
\end{equation}

As is by now well known, even though the classical action is invariant
under conformal rescalings of the metric, a trace anomaly may arise
because some counterterms may occur which fail to possess the same
invariances as the classical action (see Comments on chapter 28 in Ref.
\cite{IMPHA-114-1}). In four spacetime dimensions, such
counterterms are quadratic in Riemann, Ricci and scalar curvature
of the background. Moreover, if the boundary is nonempty, the trace
anomaly receives further contributions from local invariants which are
cubic in the extrinsic-curvature tensor $K_{ij}$ of the boundary, i.e.
$$
K_{i}^{\; j} \; K_{j}^{\; l} \; K_{l}^{i}, \;
K_{i}^{\; i}K_{lm}K^{lm}, \;
(K_{i}^{\; i})^{3}.
$$
All of this is made precise by the proportionality between trace 
anomaly and the local heat-kernel coefficient $a_{2}$ 
\cite{PHRVA-D16-3390} obtained
by studying the heat equation for an operator of Laplace type on
a Riemannian manifold with boundary \cite{CPDID-15-245, Gilkey95,
Esposito97}.

Our analysis of the energy-momentum tensor, however, was Lorentzian
rather than Euclidean. We obtained the Hadamard function as twice
the imaginary part of the Feynman Green function, and then used the
point-split method to obtain $\langle T^{\mu \nu} \rangle$. It became 
therefore important to check the consistency of the above procedure by
studying a simpler problem, and for this purpose we have here focused
on a scalar problem. Our ``Casimir'' apparatus involves a massless
scalar field in curved background, subject to homogeneous Dirichlet
conditions on parallel plates with mutual separation $a$.

Section II is a summary of basic properties of a free massless scalar field
in curved spacetime: action functional, classical energy-momentum
tensor and its regularized expression. Section III evaluates the
Feynman Green function up to first order in $\varepsilon$. Section IV
obtains the regularized and renormalized energy-momentum tensor of the
quantum theory, while Sec. V evaluates the Casimir energy. 
Discussion of the results and open problems are 
presented in Sec. VI.

\section{Free massless scalar field in curved spacetime}

The action functional for a free massles scalar field $\phi$ coupled 
to gravity reads as
\begin{equation}
S=-{1\over 2}\int \Bigr(\phi_{; \mu} \phi^{; \mu}
+\xi R \phi^{2} \Bigr) \sqrt{-g} d^{4}x,
\label{(2.1)}
\end{equation}
where $g$ is the determinant of the metric tensor $g_{\mu \nu}$, $R$
is the scalar curvature and $\xi$ is a real parameter taking the value
${1\over 6}$ in the case of conformal coupling and $0$ in the case of
minimal coupling. On requiring stationarity of the action functional
under variations of $\phi$, one obtains the field equation
\begin{equation}
(\cstok{\ } - \xi R)\phi=0,
\label{(2.2)}
\end{equation}
where $\cstok{\ } \equiv g^{\mu \nu} \nabla_{\mu} \nabla_{\nu}$ is the
wave operator.

The classical energy-momentum tensor is obtained from functional
differentiation of the classical action with respect to the metric, i.e.
\begin{equation}
T^{\mu \nu} \equiv {2\over \sqrt{-g}}
{\delta S \over \delta g_{\mu \nu}}.
\label{(2.3)}
\end{equation}
If use is made of the standard variational identities
$$
\delta g^{\mu \nu}=-g^{\mu \rho} g^{\nu \sigma} \delta g_{\rho \sigma},
\; \delta g=g \; g^{\mu \nu} \delta g_{\mu \nu},
$$
$$
\delta R=-R^{\mu \nu}\delta g_{\mu \nu}
+g^{\mu \rho} g^{\nu \sigma}(\delta g_{\rho \sigma; \mu \nu}
-\delta g_{\nu \sigma; \rho \mu}),
$$
and by introducing the anticommutator
\begin{equation}
[\phi(x),\phi(y)]_{+} \equiv \phi(x)\phi(y)+\phi(y)\phi(x),
\label{(2.4)}
\end{equation}
one finds eventually \cite{PHRVA-D14-2490}
\begin{eqnarray}
T^{\mu \nu} &=& {1\over 2}(1-2 \xi)[\phi^{;\mu},\phi^{;\nu}]_{+}
+\left(\xi-{1\over 4}\right)g^{\mu \nu}
[\phi_{;\sigma},\phi^{;\sigma}]_{+}
-\xi [\phi^{;\mu \nu},\phi]_{+} \nonumber \\
&+& \xi g^{\mu \nu} \Bigr[\phi_{; \sigma}^{\; \; \; \sigma},\phi
\Bigr]_{+}+{\xi \over 2}\left(R^{\mu \nu}-{1\over 2}g^{\mu \nu}R
\right) [\phi,\phi]_{+}.
\label{(2.5)}
\end{eqnarray}
At this stage, the point-split method comes into play, and we
re-express every anti-commutator in (2.5) according to
\begin{equation}
[\phi^{;\mu},\phi^{;\nu}]_{+}=\lim_{x' \to x}{1\over 2}
\left \{ [\phi^{;\mu'},\phi^{;\nu}]_{+}
+[\phi^{;\mu},\phi^{;\nu'}]_{+} \right \},
\label{(2.6)}
\end{equation}
\begin{equation}
[\phi^{; \mu \nu},\phi]_{+}=\lim_{x' \to x}{1\over 2}
\left \{ [\phi^{;\mu' \nu'},\phi]_{+}
+[\phi^{;\mu \nu},\phi']_{+} \right \},
\label{(2.7)}
\end{equation}
\begin{equation}
[\phi,\phi]_{+}=\lim_{x' \to x}[\phi,\phi']_{+}.
\label{(2.8)}
\end{equation}
On introducing the Hadamard two-point function (hereafter, the brackets
$\langle \; \rangle$ denote the (vacuum) expectation value)
\begin{equation}
H(x,x')= \langle [\phi(x),\phi(x')]_{+} \rangle ,
\label{(2.9)}
\end{equation}
we get (cf. \cite{PHRVA-D14-2490})
\begin{eqnarray}
\langle T^{\mu \nu} \rangle &=& \lim_{x' \to x}\biggr[
{(1-2\xi)\over 4}\Bigr(H^{;\mu' \nu}+H^{;\mu \nu'}\Bigr)
+\left(\xi -{1\over 4} \right)g^{\mu \nu}
H_{;\sigma}^{\; \; \; \sigma'} \nonumber \\
&-& {\xi \over 2} \Bigr(H^{; \mu \nu}+H^{;\mu' \nu'} \Bigr)
+{\xi \over 2}g^{\mu \nu}\Bigr(H_{;\sigma}^{\; \; \; \sigma}
+H_{;\sigma'}^{\; \; \; \sigma'} \Bigr)
+{\xi \over 2}\left(R^{\mu \nu}-{1\over 2}g^{\mu \nu}R \right)H \biggr].
\label{(2.10)}
\end{eqnarray}
Note that the use of general equations has payed off: if we now focus on
Ricci-flat spacetimes, the effect of $\xi$ remains, which would not be
obvious if Ricci-flatness were imposed in (2.1).
Hereafter, on taking the coincidence limit, we need of course the geodesic
parallel displacement bivector $P_{\; \nu'}^{\mu}$ which performs parallel
displacement of vectors along the geodesic from $x'$ to $x$. In general, it 
is defined by the differential equations \cite{PHRVA-D74-085011}
\begin{equation}
\sigma^{;\rho} P_{\; \nu';\rho}^{\mu}
=\sigma^{;\tau'}P_{\; \nu';\tau'}^{\mu}=0,
\label{(2.11)}
\end{equation}
$\sigma(x,x')$ being the Ruse--Synge world function \cite{PLMTA-32-87},
equal to half the geodesic distance between $x$ and $x'$, jointly with the
coincidence limit
\begin{equation}
\lim_{x' \to x}P_{\; \nu'}^{\mu} \equiv \Bigr[P_{\; \nu'}^{\mu}
\Bigr]=\delta_{\; \nu}^{\mu}.
\label{(2.12)}
\end{equation}
Equation (2.11) means that the covariant derivatives of $P_{\; \nu'}^{\mu}$
vanish in the directions tangent to the geodesic joining $x$ and $x'$. Thus,
the bivector $P_{\; \nu'}^{\mu}$, when acting on a vector $B^{\nu'}$
at $x'$, gives the vector ${\overline B}^{\mu}$, which is obtained by
parallel transport of $B^{\nu'}$ to $x$ along the geodesic connecting
$x$ and $x'$, i.e.
\begin{equation}
{\overline B}^{\mu}=P_{\; \nu'}^{\mu} \; B^{\nu'}.
\label{(2.13)}
\end{equation}
In our problem, we have from Eq. (1.1) the metric tensor 
\begin{equation}
g_{\mu \nu} = {\rm diag} \left(-1-\varepsilon{z\over a},1,1,1
\right),
\label{(2.14)}
\end{equation}
with contravariant form
\begin{equation}
g^{\mu \nu} \sim {\rm diag} \left(-1+\varepsilon{z\over a},1,1,1
\right)+{\rm O}(\varepsilon^{2}).
\label{(2.15)}
\end{equation}
The orthonormal tetrad $e_{\; \mu}^{a}$ relating the spacetime metric
$g_{\mu \nu}$ to the Minkowski metric $\eta_{ab}$ according to
\begin{equation}
g_{\mu \nu}=e_{\; \mu}^{a} e_{\; \nu}^{b} \eta_{ab}
\label{(2.16)}
\end{equation}
is thus given by
\begin{equation}
e_{\; \mu}^{0}=\sqrt{-g_{00}} \delta_{\; \mu}^{0}, \;
e_{\; \mu}^{i}=\delta_{\; \mu}^{i} \; \forall i=1,2,3.
\label{(2.17)}
\end{equation}
The bivector $P_{\; \nu'}^{\mu}$ hence reads as \cite{Synge60}
\begin{equation}
P_{\; \nu'}^{\mu}=g^{\mu \rho} \eta_{ab} e_{\; \rho}^{b} 
e_{\; \nu'}^{a} \sim {\rm diag} \left(
1+{\varepsilon \over 2a}(z'-z),1,1,1 \right)
+{\rm O}(\varepsilon^{2}),
\label{(2.18)}
\end{equation}
while
\begin{equation}
P_{\mu}^{\; \nu'}=g_{\mu \rho}g^{\nu' \beta'} P_{\; \beta'}^{\rho}
\sim {\rm diag} \left(1+{\varepsilon \over 2a}(z-z'),1,1,1 \right)
+{\rm O}(\varepsilon^{2}).
\label{(2.19)}
\end{equation}
The corresponding geodesic is described by the equations
\begin{equation}
t={\rm constant}, \; x_{1},x_{2}={\rm constant},
\label{(2.20)}
\end{equation}
\begin{equation}
z_{\psi} \equiv \psi(z'-z)+z, \; \psi \in [0,1].
\label{(2.21)}
\end{equation} 

\section{Feynman Green function to zeroth and first order}

The equations II were rather general, whereas now we consider,
following the introduction and our previous work
\cite{PHRVA-D74-085011, PHRVA-D76-025008}, Fermi--Walker coordinates
for a system of two parallel plates in a weak gravitational field. To
first order in the $\varepsilon$ parameter of Secs. I and II, 
the only nonvanishing Christoffel symbols associated with this metric
are therefore
\begin{equation}
\Gamma_{\; 30}^{0}=\Gamma_{\; 03}^{0}
={\varepsilon \over 2 (a+\varepsilon z)}
\sim {\varepsilon \over 2a}+{\rm O}(\varepsilon^{2}), \;
\Gamma_{\; 00}^{3} \sim {\varepsilon \over 2a}
+{\rm O}(\varepsilon^{2}).
\label{(3.1)}
\end{equation}
We are now in a position to evaluate the scalar counterpart of the 
analysis in Ref. \cite{PHRVA-D74-085011}, i.e. we compute the wave
operator $\cstok{\ }$,
the Feynman Green function of the hyperbolic operator
$(\cstok{\ }- \xi R)$, and eventually the Hadamard function and the
regularized energy-momentum tensor.

Indeed, a Green function of the operator ruling the field equation
(2.2) obeys the differential equation
\begin{equation}
(\cstok{\ }-\xi R)G(x,x')=-{\delta(x,x')\over \sqrt{-g}}.
\label{(3.2)}
\end{equation}
The Feynman Green function $G_{F}$ is the unique symmetric complex-valued
Green function which obeys the relation \cite{DeWi65}
$$
\delta G = G \; \delta F \; G,
$$
where $F$ is the invertible operator obtained from variation of the
action functional with respect to the field. This definition is well
suited for the purpose of defining the Feynman Green function even when
asymptotic flatness does not necessarily hold \cite{DeWi65}.

In our first-order expansion in the $\varepsilon$ parameter, the
scalar curvature gives vanishing contribution to Eq. (3.4), which
therefore takes the form (hereafter $\cstok{\ }^{0} \equiv
\eta^{\mu \nu}\partial_{\mu} \partial_{\nu}$)
\begin{equation}
\left(\cstok{\ }^{0}+{\varepsilon z \over (a+ \varepsilon z)}
{\partial^{2}\over \partial t^{2}}
+\Gamma_{\; 00}^{3}{a\over (a+\varepsilon z)}
{\partial \over \partial z} \right)G(x,x')
=-{\delta(x,x')\over \sqrt{-g}}.
\label{(3.3)}
\end{equation}
We now follow our work in Ref. \cite{PHRVA-D74-085011} and assume
that the Feynman Green function admits the asymptotic expansion
\begin{equation}
G_{F}(x,x') \sim G^{(0)}(x,x')+\varepsilon G^{(1)}(x,x')
+{\rm O}(\varepsilon^{2}).
\label{(3.4)}
\end{equation}
It is a nontrivial property that this asymptotics should hold
at small $\epsilon$ {\it for all} $x,x'$, no matter how close or
distant from each other are the two points. Its existence is proved
by the calculations described hereafter. Indeed,
by insertion of (3.4) into (3.3) we therefore obtain, picking out
terms of zeroth and first order in $\varepsilon$, the pair of
differential equations
\begin{equation}
\cstok{\ }^{0}G^{(0)}(x,x')=J^{(0)}(x,x'),
\label{(3.5)}
\end{equation}
\begin{equation}
\cstok{\ }^{0}G^{(1)}(x,x')=J^{(1)}(x,x'),
\label{(3.6)}
\end{equation}
having set
\begin{equation}
J^{(0)}(x,x') \equiv -\delta(x,x'),
\label{(3.7)}
\end{equation}
\begin{equation}
J^{(1)}(x,x') \equiv {z\over 2a}\delta(x,x')
-\left({z\over a}{\partial^{2}\over \partial t^{2}}
+{1\over 2a}{\partial \over \partial z} \right)G^{(0)}(x,x').
\label{(3.8)}
\end{equation}
Our boundary conditions are Dirichlet in the spatial variable $z$.
Since the full Feynman function $G_{F}(x,x')$ is required to vanish
at $z=0,a$, this implies the following homogeneous Dirichlet conditions
on the zeroth and first-order terms:
\begin{equation}
G^{(0)}(x,x') \biggr |_{z=0,a}=0,
\label{(3.9)}
\end{equation}
\begin{equation}
G^{(1)}(x,x') \biggr |_{z=0,a}=0.
\label{(3.10)}
\end{equation}
To solve Eqs. (3.5) and (3.6), we perform a Fourier analysis of 
$G^{(0)}$ and $G^{(1)}$, which remains meaningful in a weak gravitational
field \cite{PHRVA-D74-085011}, by virtue of translation invariance. 
In such an analysis we separate the
$z$ variable, i.e. we write (cf. \cite{PHRVA-D74-085011})
\begin{equation}
G^{(0)}(x,x')=\int {dk^{0}d{\vec k}_{\perp} \over (2\pi)^{3}}
\gamma^{(0)}(z,z')e^{i {\vec k}_{\perp} \cdot 
({\vec x}_{\perp}-{\vec x}_{\perp}')
-ik^{0}(x_{0}-x_{0}')},
\label{(3.11)}
\end{equation}
and similarly for $G^{(1)}(x,x')$, with a ``reduced Green function''
$\gamma^{(1)}(z,z')$ in the integrand as a counterpart of the zeroth-order
Green function $\gamma^{(0)}(z,z')$ in (3.11). Equations (3.5) and (3.6)
lead therefore to the following equations for reduced Green functions
(hereafter $\lambda \equiv \sqrt{k_{0}^{2}-k_{\perp}^{2}}$):
\begin{equation}
\left({\partial^{2}\over \partial z^{2}}+\lambda^{2} \right)
\gamma^{(0)}(z,z')=-\delta(z,z'),
\label{(3.12)}
\end{equation}
\begin{equation}
\left({\partial^{2}\over \partial z^{2}}+\lambda^{2} \right)
\gamma^{(1)}(z,z')
={z\over 2a}\delta(z,z')+\left({z\over a}k_{0}^{2}
-{1\over 2a}{\partial \over \partial z}\right)\gamma^{(0)}(z,z').
\label{(3.13)}
\end{equation}
By virtue of the Dirichlet conditions (3.11), $\gamma^{(0)}$ reads as
\begin{equation}
\gamma^{(0)}(z,z')=-{\sin(\lambda z_{<})\sin(\lambda(z_{>}-a))\over
\lambda \sin (\lambda a)},
\label{(3.14)}
\end{equation}
where $z_{<} \equiv {\rm min}(z,z')$,
$z_{>} \equiv {\rm max}(z,z')$. The evaluation of the reduced Green
function $\gamma^{(1)}$ is slightly more involved. For this purpose, we
distinguish the cases $z<z'$ and $z>z'$, and find the two equations
\begin{equation}
\left({\partial^{2}\over \partial z^{2}}+\lambda^{2} \right)
\gamma_{\pm}^{(1)}(z,z')=j_{\pm}^{(1)}(z,z'),
\label{(3.15)}
\end{equation}
where
\begin{equation}
j_{-}^{(1)}={1\over 2a}
{\lambda \cos (\lambda z)-2z k_{0}^{2}\sin (\lambda z) \over
\lambda \sin (\lambda a)} \sin (\lambda(z'-a)) \;
{\rm if} \; z< z',
\label{(3.16)}
\end{equation}
\begin{equation}
j_{+}^{(1)}={1\over 2a}{\lambda \cos (\lambda (z-a))
-2z k_{0}^{2} \sin (\lambda (z-a)) \over \lambda \sin (\lambda a)}
\sin (\lambda z') \; {\rm if} \; z > z'.
\label{(3.17)}
\end{equation}
We have therefore two different solutions in the intervals
$z < z'$ and $z > z'$. In this case the differential equation (3.13)
is solved by imposing the matching condition
\begin{equation}
\gamma_{-}^{(1)}(z',z')=\gamma_{+}^{(1)}(z',z')
\label{(3.18)}
\end{equation}
jointly with the jump condition
\begin{equation}
{\partial \over \partial z}\gamma_{+}^{(1)} \biggr |_{z=z'}
-{\partial \over \partial z}\gamma_{-}^{(1)} \biggr |_{z=z'}
={z' \over 2a}.
\label{(3.19)}
\end{equation}
Equation (3.18) is just the continuity requirement of the reduced
Green function $\gamma^{(1)}(z,z')$ at $z=z'$, 
while Eq. (3.19) can be obtained
by integrating Eq. (3.13) in a neighborhood of $z'$, since
\begin{equation}
\lim_{\epsilon \to 0}{\partial \over \partial z}\gamma^{(1)}
\biggr |_{z'-\epsilon}^{z'+\epsilon}
=\lim_{\epsilon \to 0}
\int_{z'-\epsilon}^{z'+\epsilon}{z\over 2a}\delta(z,z')dz
={z'\over 2a}.
\label{(3.20)}
\end{equation}
Bearing in mind Eq. (3.14) we can therefore write, for all $z,z'$,
\begin{eqnarray}
\gamma^{(1)}(z,z')&=& {1\over 4a \lambda^{2}} \biggr \{
\left[(k_{0}^{2}-\lambda^{2})(z+z')
-k_{0}^{2} \left(z^{2}{\partial \over \partial z}
+{z'}^{2}{\partial \over \partial z'}
\right)\right] \gamma^{(0)}(z,z')\nonumber \\
&-& k_{0}^{2}a^{2}{\sin (\lambda z) \sin (\lambda z') \over
\sin^{2}(\lambda a)} \biggr \}.
\label{(3.21)}
\end{eqnarray}

\section{Regularized and renormalized energy-momentum tensor}

In the previous section we have focused on the Feynman Green function
$G_{F}$ because it is then possible to develop a recursive scheme for
the evaluation of its asymptotic expansion at small $\varepsilon$.
However, as is clear from Eq. (2.10), we eventually need the Hadamard
function $H(x,x')$, which is obtained as \cite{PHRVA-D74-085011}
\begin{equation}
H(x,x') \equiv 2 {\rm Im}G_{F}(x,x') \sim 
2 {\rm Im}(G^{(0)}(x,x')+\varepsilon G^{(1)}(x,x'))
+{\rm O}(\varepsilon^{2}).
\label{(4.1)}
\end{equation}
The coincidence limits in (2.10), with the help of (2.13) and (2.18),
make it necessary to perform the replacements
\begin{equation}
H_{;\mu' \nu}+H_{;\mu \nu'} \rightarrow P_{\mu}^{\; \mu'}H_{;\mu' \nu}
+P_{\nu}^{\; \nu'}H_{;\mu \nu'}, \;
H_{;\sigma}^{\; \; \; \sigma'} \rightarrow g^{\sigma \rho} 
P_{\rho}^{\; \rho'} H_{;\sigma \rho'}, \;
H_{;\mu' \nu'} \rightarrow P_{\mu}^{\; \mu'} P_{\nu}^{\; \nu'}
H_{;\mu' \nu'}.
\label{(4.2)}
\end{equation}
Hence we get the asymptotic expansion at small $\varepsilon$ of the
regularized energy-momentum tensor according to (hereafter we
evaluate its covariant, rather than contravariant, form)
\begin{equation}
\langle T_{\mu \nu} \rangle \sim \langle T_{\mu \nu}^{(0)} \rangle
+\varepsilon \langle T_{\mu \nu}^{(1)} \rangle 
+{\rm O}(\varepsilon^{2}),
\label{(4.3)}
\end{equation}
where, on defining $s \equiv \pi z /a, \; s' \equiv \pi z'/a$, we find
\begin{eqnarray}
\langle T_{\mu \nu}^{(0)} \rangle &=& \left[
-{\pi^{2}\over 1440 a^{4}}-\lim_{s' \to s}
{\pi^{2}\over 2a^{4}(s-s')^{4}}\right]
\begin{pmatrix}
1 & 0 & 0 & 0 \\
0 & -1 & 0 & 0 \\
0 & 0 & -1 & 0 \\
0 & 0 & 0 & 3
\end{pmatrix}
\nonumber \\
&+& \left(\xi -{1\over 6} \right){\pi^{2} \over 8a^4}
\left[{3 -2 \sin^{2}s \over \sin^{4}s}\right]
\begin{pmatrix}
1 & 0 & 0 & 0 \\
0 & -1 & 0 & 0 \\
0 & 0 & -1 & 0 \\
0 & 0 & 0 & 0
\end{pmatrix},
\label{(4.4)}
\end{eqnarray}
and
\begin{eqnarray}
\langle T_{00}^{(1)} \rangle &=& 
{\pi \over 1440 a^{4} \sin^{4}s} \biggr[{311 \over 40}\pi
-{637 \over 40}s +{1\over 10}(43 \pi -81s)\cos 2s \nonumber \\
&+& {s -3\pi \over 40}\cos 4s +5 \sin 2s +2(\pi -s)s 
(\sin 2s -6 \cot s)\biggr] \nonumber \\
&+& \left(\xi -{1\over 6}\right){\pi \over 48 a^{4}\sin^{4}s}
\Bigr[2(\pi+s)(2+\cos 2s)+{5\over 2} \sin 2s \nonumber \\
&+& (\pi-s)s (\sin 2s -6 \cot s)\Bigr]
-\lim_{s' \to s}{\pi s \over 2a^{4}(s-s')^{4}},
\label{(4.5)}
\end{eqnarray}
\begin{eqnarray}
\langle T_{11}^{(1)} \rangle &=& {\pi \over 7200 a^{4}}
\biggr[\pi-2s +{5\over \sin^{2}s} \Bigr(2(\pi-2s)
\left(-2+{3\over \sin^{2}s}\right) \nonumber \\
&+& \cot s \left(5+2(\pi -s)s -6(\pi -s){s\over \sin^{2}s}\right)
\Bigr) \biggr] \nonumber \\
&+& \left( \xi-{1\over 6} \right) {\pi \over 96a^{4} \sin^{5}s}
\Bigr[(11(\pi-s)s-1)\cos s \nonumber \\
&+& ((\pi-s)s+1)\cos 3s-2(\pi-2s)(3 \sin s +\sin 3s)\Bigr],
\label{(4.6)}
\end{eqnarray}
\begin{equation}
\langle T_{22}^{(1)} \rangle = \langle T_{11}^{(1)} \rangle ,
\label{(4.7)}
\end{equation}
\begin{equation}
\langle T_{33}^{(1)} \rangle=-{\pi^{2}\over 1440 a^{4}}
+{\pi s \over 720 a^{4}} +\left(\xi -{1\over 6}\right)
{\pi \over 16 a^{4}}{\cos s \over \sin^{3} s}.
\label{(4.8)}
\end{equation}

The next step of our analysis is the renormalization of the regularized
energy-momentum tensor. For this purpose, following our work in
Ref. \cite{PHRVA-D74-085011}, we subtract the energy-momentum tensor 
evaluated in the absence of bounding plates, i.e.
\begin{equation}
\langle {\widetilde T}_{\mu \nu}^{(0)} \rangle 
=-\lim_{s' \to s}{\pi^{2}\over 2a^{4}(s-s')^{4}}
\begin{pmatrix}
1 & 0 & 0 & 0 \\
0 & -1 & 0 & 0 \\
0 & 0 & -1 & 0 \\
0 & 0 & 0 & 3
\end{pmatrix},
\label{(4.9)}
\end{equation}
and
\begin{equation}
\langle {\widetilde T}_{\mu \nu}^{(1)} \rangle 
=-\lim_{s' \to s}{\pi s \over 2a^{4}(s-s')^{4}}
\begin{pmatrix}
1 & 0 & 0 & 0 \\
0 & 0 & 0 & 0 \\
0 & 0 & 0 & 0 \\
0 & 0 & 0 & 0
\end{pmatrix}.
\label{(4.10)}
\end{equation}
To test consistency of our results we should now check whether our
regularized and renormalized energy-momentum tensor is covariantly 
conserved, since otherwise we would be outside the realm of quantum
field theory in curved spacetime, which would be unacceptable. Indeed,
the condition
\begin{equation}
\nabla^{\mu} \langle T_{\mu \nu} \rangle =0
\label{(4.11)}
\end{equation}
yields, working up to first order in $\varepsilon$, the pair of equations
\begin{equation}
{\partial \over \partial z} \langle T_{33}^{(0)} \rangle=0,
\; (\varepsilon^{0} \; {\rm term})
\label{(4.12)}
\end{equation}
\begin{equation}
{\partial \over \partial z} \langle T_{33}^{(1)} \rangle
+{1\over 2a}\Bigr(\langle T_{00}^{(0)} \rangle 
+ \langle T_{33}^{(0)} \rangle \Bigr)=0 \;
(\varepsilon^{1} \; {\rm term}),
\label{(4.13)}
\end{equation}
which are found to hold identically for all values of $\xi$ 
in our problem.

The trace of $\langle T_{\mu \nu} \rangle$ is obtained as
($g^{\mu \nu}$ being the contravariant metric in (2.15))
\begin{equation}
\tau \equiv g^{\mu \nu} \langle T_{\mu \nu} \rangle
\sim \eta^{\mu \nu} \langle T_{\mu \nu}^{(0)} \rangle
+\varepsilon \Bigr[\eta^{\mu \nu} \langle T_{\mu \nu}^{(1)}
\rangle +{z\over a} \langle T_{00}^{(0)} \rangle \Bigr]
+{\rm O}(\varepsilon^{2}),
\label{(4.14)}
\end{equation}
from which we find a $\xi$-dependent part
\begin{eqnarray}
\tau_{\xi}&=& \left(\xi-{1\over 6}\right)\biggr \{ 
-{3\pi^{2}(2+\cos 2s)\over 8a^{4}\sin^{4}s}
-\varepsilon {\pi \over 32 a^{4}\sin^{5}s}
\Bigr[(1-11(\pi -s)s)\cos s \nonumber \\
&-& (1+(\pi-s)s)\cos 3s +2(\pi-2s)(3 \sin s + \sin 3s)\Bigr]
\biggr \}.
\label{(4.15)}
\end{eqnarray}
Interestingly, the value $\xi={1\over 6}$ which yields conformal
invariance of the classical action (2.1) with wave equation (2.2)
is the same as the value of $\xi$ yielding no trace anomaly, unlike
what happens for Maxwell theory in Ref. \cite{PHRVA-D74-085011},
where the conformally invariant action is found to lead to a trace 
anomaly in quantum theory with mixed boundary conditions.

\section{Casimir energy and pressure}

In order to evaluate the energy density $\rho$ of our ``scalar'' Casimir 
apparatus, we project the regularized and renormalized
energy-momentum tensor along a unit timelike vector
$u^{\mu}=\left(-{1\over \sqrt{-g_{00}}},0,0,0 \right)$. This yields
\begin{eqnarray}
\rho &=& \langle T_{\mu \nu} \rangle u^{\mu} u^{\nu}
=-{\pi^{2}\over 1440 a^{4}}+{\pi \over 7200 a^{4}}
\biggr[-3 \pi+6s +{10\over \sin^{2}s}\Bigr(2(\pi-2s) \nonumber \\
& \times & \left(-2+{3\over \sin^{2}s}\right)
+\cot s \left((5+2(\pi-s)s+6{s(-\pi+s)\over 
\sin^{2}s}\right)\Bigr)\biggr]
\varepsilon \nonumber \\
&+& \left(\xi-{1\over 6}\right) \biggr \{ 
{\pi^{2}(2+\cos 2s)\over 8a^{4} \sin^{4}s}
-{\pi \over 192 a^{4} \sin^{5}s}\biggr[\Bigr(-5+22(\pi-s)s\Bigr)
\cos s \nonumber \\
&+& \Bigr(5+2(\pi-s)s \Bigr)\cos 3s -4(\pi-2s)(3 \sin s +\sin 3s)
\biggr] \varepsilon \biggr \}.
\label{(5.1)}
\end{eqnarray}
The energy $E$ stored within our Casimir cavity is given by
\begin{equation}
E=\int_{V_{c}}d^{3}\Sigma \sqrt{-g}\rho,
\label{(5.2)}
\end{equation}
where $d^{3}\Sigma$ is the volume element of an observer with
four-velocity $u^{\mu}$, and $V_{c}$ is the volume of the cavity.
The integration in (5.2) requires the use of approximating domains, 
i.e. the $z$-integration is performed in the interval
$(\zeta,a-\zeta)$, corresponding to ${\pi \over a}(\zeta,a-\zeta)$ 
in the $s$ variable, taking eventually
the $\zeta \to 0$ limit. We thus obtain
\begin{equation}
E_{\xi}=-{\pi^{2}A\over 1440 a^{3}}
-{\pi^{2}A \varepsilon \over 5760 a^{3}}
+\left(\xi -{1\over 6}\right){\pi A \over 4a^{3}}
\left(1+{\varepsilon \over 4}\right)
\lim_{\zeta \to 0}{\cos \zeta \over \sin^{3}\zeta},
\label{(5.3)}
\end{equation}
where $A$ is the area of parallel plates. Note that the conformal coupling
value $\xi={1\over 6}$ is picked out as the only value of $\xi$ for which
the Casimir energy remains finite. In this case, reintroducing the constants
$\hbar,c$ and writing explicitly $\varepsilon$, we find
\begin{equation}
E_{c}=-{\hbar c \pi^{2}\over 1440}{A \over a^{3}}
\left(1+{1\over 2}{ga \over c^{2}}\right).
\label{(5.4)}
\end{equation}

In the same way, the pressure $P_{\xi}$ on the parallel plates is
found to be \cite{PHRVA-D74-085011}
\begin{equation}
P_{\xi}(z=0)={\pi^{2}\over 480 a^{4}}
+{\pi^{2}\varepsilon \over 1440 a^{4}}
-\left(\xi-{1\over 6}\right){\pi \varepsilon \over 16a^{4}}
\lim_{s \to 0}{\cos s \over \sin^{3}s},
\label{(5.5)}
\end{equation}
\begin{equation}
P_{\xi}(z=a)=-{\pi^{2}\over 480 a^{4}}
+{\pi^{2}\varepsilon \over 1440 a^{4}}
+\left(\xi-{1\over 6}\right){\pi \varepsilon \over 16a^{4}}
\lim_{s \to \pi}{\cos s \over \sin^{3}s}.
\label{(5.6)}
\end{equation}
Once again, one can get rid of divergent terms by setting 
$\xi={1\over 6}$, which leads to
\begin{equation}
P_{c}(z=0)={\pi^{2}\over 480} {\hbar c \over a^{4}}
\left(1+{2\over 3}{ga \over c^{2}}\right),
\label{(5.7)}
\end{equation}
\begin{equation}
P_{c}(z=a)=-{\pi^{2}\over 480}{\hbar c \over a^{4}}
\left(1-{2\over 3}{ga \over c^{2}}\right).
\label{(5.8)}
\end{equation}

\section{Concluding remarks}

In our paper we have obtained, for the first time in the literature,
a detailed evaluation of the regularized and renormalized energy-momentum
tensor for a scalar Casimir apparatus in a weak gravitational field,
the plates being parallel plates upon which a real massless scalar
field is required to obey homogeneous Dirichlet conditions. Moreover,
the trace anomaly is found to vanish provided the free scalar field
is conformally coupled to gravity. The conformal coupling also ensures
a finite Casimir energy and finite pressure on the plates.

The electromagnetic case is definitely more involved: the boundary 
conditions are a mixture of Dirichlet and Robin conditions, and if the
Casimir apparatus is set in a weak gravitational field the trace anomaly
is not found to vanish \cite{PHRVA-D74-085011}, despite 
that the classical Maxwell action is conformally invariant in four
spacetime dimensions. At least three investigations are now in order:
\vskip 0.3cm
\noindent
(i) To repeat the scalar analysis with Robin boundary conditions,
to understand whether the latter are responsible for the nonvanishing
trace anomaly found in Ref. \cite{PHRVA-D74-085011} for the
electromagnetic case.
\vskip 0.3cm
\noindent
(ii) To work out the relation (if any) between our small-$\varepsilon$
asymptotics of the Feynman Green function $G_{F}$, with the following
limit as $x' \rightarrow x$, and the usual approach where one first
considers the Schwinger--DeWitt asymptotics of $G_{F}$ at small values
of the world function 
\cite{PHRVA-D14-2490, PHRVA-D17-946, CQGRD-8-603}. 
The two approaches are not obviously equivalent nor easily comparable,
but the novel features found in Ref. \cite{PHRVA-D74-085011} make it
compelling to produce further efforts along both lines.
\vskip 0.3cm
\noindent
(iii) To relate our results to the enlightening energy-momentum analysis in
Ref. \cite{PRSLA-A354-79}. 

\acknowledgments
Conversations with G. Bimonte, E. Calloni, S. Fulling and K. Milton
have been very helpful.
G. Esposito is grateful to the Dipartimento di Scienze Fisiche of
Federico II University, Naples, for its hospitality and support.
The work of L. Rosa has been partially supported by PRIN 
{\it FISICA ASTROPARTICELLARE}.


\begin{references}
\bibitem{KNAWA-51-793}
H.B.G. Casimir, Proc. K. Ned. Akad. Wet. Rev. {\bf 51}, 793 (1948).
\bibitem{PRPLC-353-1}
M. Bordag, U. Mohideen, and V.M. Mostepanenko, Phys. Rep. {\bf 353},
1 (2001).
\bibitem{JPAGB-37-R209}
K. Milton, J. Phys. A {\bf 37}, R209 (2004).
\bibitem{RNCIB-27N6-1}
V.V. Nesterenko, G. Lambiase, and G. Scarpetta, Riv. Nuovo Cimento,
Ser. 4, {\bf 27}, Issue 6, 1 (2004).
\bibitem{RPPHA-68-201}
S.K. Lamoreaux, Rep. Prog. Phys. {\bf 68}, 201 (2005).
\bibitem{IJSQE-13-400}
F. Capasso, J.N. Munday, D. Iannuzzi, and H.B. Chan, IEEE J. Sel. Top.
Quantum Electron. {\bf 13}, 400 (2007).
\bibitem{PHRVA-D76-025004}
S.A. Fulling, K.A. Milton, P. Parashar, A. Romeo, K.V. Shajesh,
and J. Wagner, Phys. Rev. D {\bf 76}, 025004 (2007).
\bibitem{JPAGB-40-10935}
K.A. Milton, P. Parashar, K.V. Shajesh, and J. Wagner, 
J. Phys. A {\bf 40}, 10935 (2007).
\bibitem{PHRVA-D74-085011}
G. Bimonte, E. Calloni, G. Esposito, and L. Rosa, Phys. Rev.
D {\bf 74}, 085011 (2006); erratum ibid. D {\bf 75}, 049904 (2007);
erratum ibid. D {\bf 75}, 089901 (2007).
\bibitem{PHRVA-D76-025008}
G. Bimonte, E. Calloni, G. Esposito, and L. Rosa, Phys. Rev.
D {\bf 76}, 025008 (2007).
\bibitem{IMPHA-114-1}
B.S. DeWitt, {\it The Global Approach to Quantum Field Theory},
International Series of Monographs on Physics {\bf 114}
(Clarendon Press, Oxford, 2003).
\bibitem{PHRVA-D16-3390}
J.S. Dowker and R. Critchley, Phys. Rev. D {\bf 16}, 3390 (1977).
\bibitem{CPDID-15-245}
T.P. Branson and P.B. Gilkey, Commun. Part. Diff. Eqs. {\bf 15},
245 (1990).
\bibitem{Gilkey95}
P.B. Gilkey, {\it Invariance Theory, The Heat Equation and The 
Atiyah--Singer Index Theorem} (CRC Press, Boca Raton, 1995).
\bibitem{Esposito97}
G. Esposito, A.Yu. Kamenshchik, and G. Pollifrone, {\it Euclidean
Quantum Gravity on Manifolds With Boundary}, Fundam. Theor. Phys. 
{\bf 85} (Kluwer, Dordrecht, 1997).
\bibitem{PHRVA-D14-2490}
S.M. Christensen, Phys. Rev. D {\bf 14}, 2490 (1976).
\bibitem{PLMTA-32-87}
H.S. Ruse, Proc. London Math. Soc. {\bf 32}, 87 (1931);
J.L. Synge, Proc. London Math. Soc. {\bf 32}, 241 (1931).
\bibitem{Synge60}
J.L. Synge, {\it Relativity: The General Theory}
(North--Holland, Amsterdam, 1960).
\bibitem{DeWi65}
B.S. DeWitt, {\it Dynamical Theory of Groups and Fields}
(Gordon \& Breach, New York, 1965).
\bibitem{PHRVA-D17-946}
S.M. Christensen, Phys. Rev. D {\bf 17}, 946 (1978).
\bibitem{CQGRD-8-603}
D.M. McAvity and H. Osborn, Class. Quantum Grav. {\bf 8}, 603 (1991).
\bibitem{PRSLA-A354-79}
P. Candelas and D. Deutsch, Proc. Roy. Soc. Lond. A {\bf 354}, 79 (1977).
\end{references}
\end{document}